\documentclass{article}

\usepackage{amsmath}
\usepackage{amsthm}
\usepackage{amssymb}
\usepackage{amsfonts}
\usepackage{multirow}
\usepackage{dsfont}
\usepackage{array}
\usepackage{xcolor}
\usepackage{graphicx}
\usepackage{subcaption}
\usepackage{hyperref}
\usepackage{fullpage}

\graphicspath{{Figures/}}
\usepackage[utf8]{inputenc} 

\newcommand{\tup}[1]{\left< #1 \right>}	
\newcommand{\rem}[1]{}
\newcommand{\sub}{\subseteq} 

\begin{document}

\title{Hypergraph Models of Biological Networks to Identify Genes Critical to Pathogenic Viral Response}


\author{Song Feng${}^1$, Emily Heath${}^2$, Brett Jefferson${}^3$, Cliff Joslyn${}^{3, 4}$, Henry Kvinge${}^3$, \and
Hugh D. Mitchell${}^1$, Brenda Praggastis${}^3$, Amie J. Eisfeld${}^5$, Amy C. Sims${}^{6}$, \and
Larissa B. Thackray${}^7$, Shufang Fan${}^5$, Kevin B. Walters${}^5$, Peter J. Halfmann${}^5$, \and
Danielle Westhoff-Smith${}^5$, Qing Tan${}^7$, Vineet D. Menachery${}^{8,9}$, Timothy P. Sheahan${}^8$, \and
Adam S. Cockrell${}^{10}$, Jacob F. Kocher${}^8$, Kelly G. Stratton${}^3$, Natalie C. Heller${}^3$, \and
Lisa M. Bramer${}^3$, Michael S. Diamond${}^{7,11,12}$, Ralph S. Baric${}^8$, Katrina M. Waters${}^{1,13}$, \and
Yoshihiro Kawaoka${}^{5,14,15,16}$, Jason E. McDermott${}^{1,17}$, Emilie Purvine${}^{3,}$\thanks{Emilie.Purvine@pnnl.gov}}


\maketitle

\begin{abstract} 
\noindent\textbf{Background:} Representing biological networks as graphs is a powerful approach to reveal underlying patterns, signatures, and critical components from high-throughput biomolecular data. However, graphs do not natively capture the multi-way relationships present among genes and proteins in biological systems. Hypergraphs are generalizations of graphs that naturally model multi-way relationships and have shown promise in modeling systems such as protein complexes and metabolic reactions. In this paper we seek to understand how hypergraphs can more faithfully identify, and potentially predict, important genes based on complex relationships inferred from genomic expression data sets.

\vspace{0.5em}
\noindent\textbf{Results:} We compiled a novel data set of transcriptional host response to pathogenic viral infections and formulated relationships between genes as a hypergraph where hyperedges represent significantly perturbed genes, and vertices represent individual biological samples with specific experimental conditions. We find that hypergraph betweenness centrality is a superior method for identification of genes important to viral response when compared with graph centrality.

\vspace{0.5em}
\noindent\textbf{Conclusions:} Our results demonstrate the utility of using hypergraphs to represent complex biological systems and highlight central important responses in common to a variety of highly pathogenic viruses.

\vspace{0.5em}
\noindent\textbf{Keywords:} Systems biology, Hypergraph, Viral infection, Biological networks, SARS, MERS, Influenza, West Nile Virus, Host response, Viral pathogenesis
\end{abstract}

\section{Background}

Identifying molecular signatures critical to a biological process requires an accurate model of both the process and the biological system in which it occurs.
Thus it is essential that such a model be able to represent its target with complexity commensurate with that of the system itself, rather than presenting only a simplified view.
Commonly, biological systems and processes present as complex networks of interacting entities, for example within and between genes, pathways, and complexes.
Graphs are frequently used to model these interactions, but since graphs can only capture interactions between pairs of entities, they fall short in many cases and are not able to model the full complexity present in biological systems and processes.

In this paper we investigate the role that hypergraph models, as mathematical generalizations of graph models, can play in providing the necessary complexity to capture multi-way interactions in biological systems inferred from genomic expression data.
We assert that the complexity provided by hypergraphs more closely represents the systems being studied.
In order to validate this assertion we provide a comparison between the use of graph and hypergraph centrality metrics to identify genes that are critical in host responses to viral infection.
Our findings show that the genes identified using a hypergraph model align better with genes previously known to correlate with viral response than do genes identified using similar metrics from graphs or using average fold change for each gene across all experimental conditions.

\subsection{Network science for high-throughput data analysis}
Modern biology has been transformed by the rapid growth of technologies to measure the abundance of large numbers of biological entities over many samples simultaneously. Such high-throughput methods like transcriptomics, proteomics, metabolomics, and lipidomics allow researchers to gain unparalleled scientific insight into the mechanisms underlying biological systems.
A wide range of biological questions have been addressed using such systems biology approaches including questions related to cancer, microbiomes, and infectious disease.
Analysis methods for high-throughput measurements are also varied, ranging from simple statistical tests for differential abundance (between control and experimental conditions, for example), to dimensionality reduction, to machine learning, all with the aim of extracting more relevant information from the high-dimensional and often noisy measurements.

A powerful approach for modeling systems using high-throughput data is network biology.
Here biological systems are modeled as graphs, with molecular entities (genes, proteins, metabolites) represented as vertices, and relationships between molecules represented as edges connecting them.
Relationships between molecules are generally determined from existing knowledge of protein-protein interactions, regulatory interactions, metabolic networks, or can be inferred from high-throughput systems biology data.
We and others have used networks inferred from correlation or mutual information between abundance profiles of genes and proteins to identify critical entities \cite{mcdermott2016effect,mitchell2019,tran2020}, integrate different data types \cite{adourian2008,diamond2010,maier2017,mcclure2019}, and represent and predict temporal dynamics in the system \cite{lempp2019,mcdermott2012,mcdermott2011}.

\subsection{Hypergraphs for complex network models}\label{sec:intro_hg}
While graph-based methods have been quite successful in the biological domain, their ability to model complex relationships amongst entities is necessarily limited. Graphs inherently model relationships (edges) between {\em pairs} of entities (vertices).
But biological systems are replete with  relationships among {\em many} entities, for example in  protein complexes, transcription factor and microRNA regulation networks, lipid and metabolite enzyme-substrate interactions, metabolic networks, pathways, and protein function annotations.
Relationships may be interactions, for example, metabolites working together in a metabolic process, or they may represent some commonality among the entities, like genes being differentially expressed in the same conditions, or regulated by the same transcription factor.
In a graph model all of these multi-way relationships would be represented as groups of pairs of subunits, which would not fully capture how groups of components interact or have similar behavior.

Sometimes sets of related components are already understood, and sometimes they need to be discovered in experimental data, like high-throughput `omics.
In either event, a higher order mathematical model is needed.
The mathematical object that {\em natively} represents multi-way interactions amongst entities is called a  ``hypergraph''.
In contrast to a graph, in a hypergraph the relationships amongst entities (still called vertices) are connected generally by ``hyperedges'', where each hyperedge is an arbitrary subset of vertices.
Thus every graph is a hypergraph in which each hyperedge happens to have exactly two vertices.
A challenge for scientists is to  recognize the presence of hypergraph structure in their data, and to judge the relative value of representing them natively as hypergraphs or reducing them to graph structures.
Mathematically, hypergraphs  are closely related to ``bipartite graphs'', those with two distinct sets of nodes so that there are connections only between the two sets, and not within each set.
In a bipartite graph corresponding to a given hypergraph one group of nodes plays the role of hypergraph vertices, and the other of the hyperedges themselves.
While this observation can be helpful in the design of scalable hypergraph algorithms it falls outside the scope of this paper.

\begin{figure}[h]
    \centering
    \includegraphics[width=0.9\textwidth]{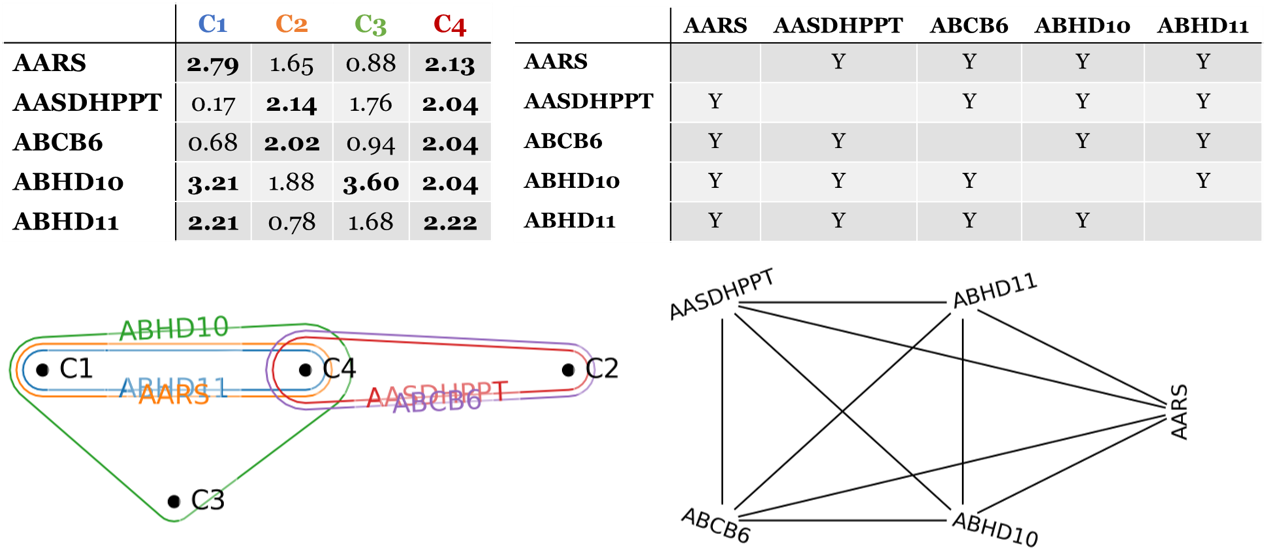}
    \caption{Transcriptomics example comparing graphs and hypergraphs. (Upper left) $\log_2$-fold change values for 5 genes across 4 conditions. (Lower left) Visualization of a corresponding hypergraph. (Upper right) Adjacency matrix for expression data. (Lower right) Underlying graph.}
    \label{fig:omics_hg_graph}
\end{figure}

Hypergraph models allow for higher fidelity representation of data that may contain multi-way relationships, albeit at the price of a higher complexity model.
An example using a small subset of transcriptomic expression data is shown in Figure~\ref{fig:omics_hg_graph}.
In the upper left is an expression matrix with $\log_2$-fold change values for five genes (rows) across four experimental conditions (columns).
The lower left shows a hypergraph representation of the data, with
each gene modeled as a hyperedge surrounding those conditions (vertices) for which the $\log_2$-fold change is greater than 2. Those cells in the expression matrix are shown in bold, distinguishing those conditions which are included in that gene's hyperedge.
The upper right of Figure~\ref{fig:omics_hg_graph} shows a matrix produced from one possible graph-based approach to representing these data. Here each pair of genes is related if there is at least one condition for which both genes have $\log_2$-fold change greater than 2.
This would then be  interpreted as an ``adjacency matrix'' of a graph, which is then shown in the lower right.
It can be seen that this graph representation necessarily loses a great deal of information, boiling down the rich interaction structure that we know to be present to a fully connected graph on all five genes.
For example, the hypergraph shows that two pairs of genes---AARS and ABHD11, AASDHPPT and ABCB6---are much more related than other pairs.
This fact is not apparent in the graph model.

Although graphs and graph theory dominate network science applications and methods \cite{barabasi2016network}, hypergraphs are well-known objects in mathematics and computer science. They have a history of use in a range of applications \cite{IaIPeG19,klamt2009hypergraphs,PaAPeG17}, and are seeing increasingly wide adoption \cite{adourian2008,JaMLuL18,JoCAkS20,LeWReG19,MiM00}.
In the biological literature we have seen hypergraphs used to model gene and protein interaction networks, pathways, and metabolic networks as derived from a variety of data types.
In many of these cases the authors derive hypergraphs from an underlying graph, rather than directly from data.
For example, Chitra built a hypergraph model based on an existing graph model of gene interaction networks \cite{randomwalks}.
They adapt the PageRank algorithm to hypergraphs in order to study disease-gene prioritization, and find that for monogenic diseases hypergraph PageRank noticeably outperforms graph PageRank.
Tran studies protein function prediction building a graph from a similarity matrix derived from gene expression data \cite{tran2012hypergraph} and then applying soft clustering to this graph to produce a hypergraph.
Function prediction using this hypergraph is then shown to be superior to predictions based on graphs.
Protein-protein interaction networks are studied by Klamt {\it et al.} using graph algorithms to find sets of independent elements or tightly connected elements \cite{klamt2009hypergraphs}.
In those three papers the authors infer a hypergraph from a graph structure rather than directly from data.

Biological researchers may be familiar with the \emph{directed} variety of hypergraph (e.g., used by Klamt \cite{klamt2009hypergraphs}). These hypergraphs can represent ``flows'' in that each hyperedge is partitioned into two disjoint groups of vertices such that the first group are understood as inputs and the second as outputs.
These and other issues in hypergraph theory and hypernetwork science can have great significance for general complex systems modeling, including biological systems, but will fall outside the scope of this paper.

Ramadan {\it et al.} use hypergraphs to model the yeast proteome, where proteins are vertices and complexes are hyperedges \cite{hypergraph_yeast}, and apply an algorithm that finds tightly connected vertices to identify the core proteome.
Finally, Zhou and Nakhleh study the claim that metabolic networks are hierarchical and small-world \cite{hypergraph_metabolic}.
While this claim comes from a graph model of the networks, Zhou and Nakleh instead model the metabolic networks of \emph{E. coli} as a hypergraph and show that the claimed hierarchy and scaling properties are not supported.
This result in particular conveys a critical message: when biological interactions are simplified into pairwise relationships and modeled using a graph, they can exhibit very different structure than when their true complexity is modeled using a hypergraph.
Because of this structural variance, conclusions drawn based on the graph could provide misleading results.
Although the data we consider are different, our method is similar to these last two papers in that we build hypergraphs directly from biological data rather than inferring a hypergraph from a standard graph model of the data.
We have not observed researchers building hypergraphs directly from `omics data, as we will in this paper.

\subsection{Modeling host response in viral pathogenesis}

Viral infection causes a response in the host cells in which the expression of a variety of cell systems are up- or down-regulated.
The pathogenesis of the infection is reflected in the signature of host responses elicited by each virus.
Host response to viral infection has been extensively studied for decades, yet the root mechanisms of why some infections are severe and some are not remain poorly understood.
However, high-throughput molecular approaches offer a way to discover novel host response genes, proteins, and pathways that contribute to the systems-level development of pathogenesis.
A major advantage of such a systems biology approach to pathobiology is the ability to identify novel, key elements of a biological process, such as which regulators are involved in critical processes.
High-throughput profiling methods (e.g. transcriptomics) provide powerful tools for examining how entire systems respond to different perturbations such as acute disease.
Network reconstruction provides the opportunity to utilize all available data and is a critically important tool for representing complex sets of interactions \cite{desmetNetworks}.

In this paper we build and explore a \emph{hypergraph} model of host response using transcriptomics data from viral infection by five highly pathogenic viruses in a number of biological systems.
We found that gene rankings computed using hypergraph centrality were highly enriched for known immune and infection-related genes.
While rankings derived from graphs constructed using other traditional computational biology techniques applied to the same infection data also resulted in rankings enriched for critical genes, we demonstrate that hypergraph-based metrics yield superior enrichment results.
These results highlight the usefulness of hypergraphs for exploring mechanisms of virus pathobiology.

\section{Methods}

\subsection{Data acquisition and processing}
Microarray datasets collected from 2014 to 2017 and available from the Gene Expression Omnibus (GEO) were gathered and processed. GEO accession IDs: GSE80059, GSE86533, GSE69027, GSE76600, GSE80697, GSE69945, GSE68945, GSE72008, GSE65575, GSE79458, GSE86528, GSE100496, GSE81909, GSE86530, GSE100504, GSE106523, GSE86529, GSE100509, GSE108594, GSE77193, GSE77160, GSE78888, GSE33267, GSE37827, GSE48142, GSE33266, GSE49262. 
While details of experimental systems and conditions can be gathered from individual accessions from GEO, we list the infection conditions here:

\begin{description}

\item[Ebola Virus:] (Wild type and two mutants) in human hepatocyte cells. 
\item[Influenza Virus:]  H7N9 (Wild type and two mutants) in human lung epithelial cells, H1N1 (wild type) in human lung epithelial cells, H1N1 (wild type) in mouse lung, H5N1 (wild type and one mutant) in mouse lung, H7N9 (wild type and two mutants) in mouse lung. 
\item[MERS-coronavirus:] (Wild type and four mutants) in human lung epithelial cells, (wild type only) in \emph{ex vivo} human epithelial cells, in \emph{ex vivo} human lung fibroblasts, and \emph{ex vivo} human lung microvascular endothelial cells. 
\item[SARS-coronavirus:] (Wild type and four mutants) in human lung epithelial cells and mouse lung. 
\item[West Nile Virus:] (Wild type and one mutant) in mouse cerebral cortex, mouse cerebellum and mouse lymph node. 
\end{description}
Conversion from mouse to human gene symbols was done by capitalizing the mouse symbols. Quality control, background correction, quantile normalization, and differential expression analysis were performed in R using the limma package \cite{ritchie2015limma}. Replicates from each infection condition were compared to corresponding time-matched mock-infected samples to obtain $\log_2$-fold change values, as well as adjusted $p$-values.

\subsection{Hypergraph representations and centrality metrics}
Formally, a {\bf hypergraph} is a structure $\tup{V,E}$, with $V=\{v_j\}_{j=1}^n$ a  set of vertices, and $E = \{e_i\}_{i=1}^m$ a family of hyperedges with each $e_i \sub V$.
Hyperedges can come in different sizes, $|e_i|$, possibly ranging from the singleton $\{v\} \sub V$ (distinct from the element $v \in V$) to the entire vertex set $V$.
A hyperedge $e=\{v_1,v_2\}$ where $|e|=2$ is the same as a graph edge and so it follows that all graphs are hypergraphs, specifically identified as being ``2-uniform''.
In the remainder of the paper where clear from context we may use the terms edge and hyperedge interchangeably.

We construct a hypergraph from transcriptomics data using a threshold approach, much like the example in Figure \ref{fig:omics_hg_graph}.
Again, vertices $v_j$ will represent individual biological or experimental ``conditions'' (e.g., mouse lung cells treated with a strain of Influenza virus and sampled at 8 hours) and hyperedges $e_i$ represent genes. Thus for us, a hyperedge $e_i$ is a gene $i$ that includes a collection of conditions $j$ as its vertices $v_j$.
For each condition, we transform the $\log_2$-fold change values (relative to uninfected mock) for all of the genes into absolute value $z$-scores. 
Then, the vertex representing condition $X$ is contained in the hyperedge representing gene $G$ if the absolute value $z$-score for $G$ in $X$ is greater than or equal to 2 and the adjusted $p$-value for that $\log_2$-fold change measurement is less than 0.05. Since transcriptomics $\log_2$-fold change values tend to be normally distributed for each condition across all genes a $z$-score transformation is a reasonable way to get all conditions onto the same scale before applying a threshold. The specific thresholds on $z$-score and $p$-value were chosen as commonly used in the field, and in exploring other values we have verified limited sensitivity to them.

In this way, hyperedges correspond to genes, and indicate the groups of conditions in which that gene is both highly perturbed (either up or down) from the mock infected control condition, and for which that perturbation is statistically significant. 
In the remainder of this paper we will say that the gene is ``significantly perturbed'' in the condition.
Unlike hypergraph models of pathways or metabolic reactions, hypergraphs constructed from high-throughput data do not necessarily represent actual biological interactions but rather capture relationships based on similar behavior among entities. 

It is important to point out that this method to construct a hypergraph using thresholds on absolute value of $z$-score and $p$-value is a specific case of a flexible framework for how hyperedges can be formed from `omics data.
Applying other thresholds will result in different hypergraph models of the same data, to potentially answer different questions.
For example, in order to understand the relationship and behavioral similarity among up-regulated genes one might consider a gene hyperedge to contain those conditions for which the gene has high raw (as opposed to absolute value) $z$-score or $\log_2$-fold change, as in the Figure \ref{fig:omics_hg_graph} example.
One could also form edges from conditions for which a gene has a highly negative $z$-score or fold change, to explore the structure of down-regulated genes. 
We chose a threshold on the absolute value of $z$-score in this paper as an attempt to understand genes which are perturbed at all in response viral infection. 

We recognize that this formulation is fundamentally different from a typical graph approach to systems biology data. One such example of a graph approach is context likelihood of relatedness (CLR) in which genes that show similar expression patterns across \emph{all} conditions, as measured by mutual information, are linked together \cite{faith2007large}. 
Our approach to constructing hypergraphs from the data can be seen as having greater sensitivity and flexibility since it allows similarity between genes to be assessed across any number of conditions rather than requiring assessment across all conditions. 

Another difference between our hypergraph formulation and typical graph approaches is that in graph approaches vertices represent genes and edges indicate some relationship between genes such as interaction or expression correlation.
Our motivation for swapping the roles of vertices and edges is for the sake of clarity in our description of hypergraph centrality measures below.
Moreover, as a technical matter, each hypergraph $H$ determines another one, called its   ``dual''  $H^*$, formed exactly by swapping the roles of vertices and edges \cite{berge1984hypergraphs}.
Therefore, the dual to our hypergraph formulation has the more traditional form with genes as vertices, but the description of hypergraph centrality in this setting would be less intuitive. 

As in graphs, the way in which hyperedges connect vertices in complex patterns is central to the study of hypergraphs.
Our focus in this paper is applying generalizations of graph centrality measures to hypergraphs built from transcriptomics data to identify important genes.
In order to define these hypergraph centrality measures we must first introduce the notions of a hypergraph walk and distance \cite{aksoy2019hypernetwork}.
Given two hyperedges $e, f \in E$, an {\bf $s$-walk of length $k$} between $e$ and $f$ is a sequence of hyperedges $e_0, e_1, \ldots, e_k$ such that $e_0 = e$, $e_k = f$, and $s \le |e_i \cap e_{i+1}|$ for all $0 \leq i \leq k-1$. 
In other words, an $s$-walk is a sequence of edges such that pairwise intersections between neighboring edges have size at least $s$.
Note that a graph walk is a $1$-walk. 
We note that one could define a hypergraph $s$-walk on to be between vertices rather than hyperedges, as is typically done in a graph.
But as above, for the sake of clarity in defining centrality measures we use this edge-based definition.


Continuing to follow  Aksoy {\it et al.} \cite{aksoy2019hypernetwork}, for a fixed $s > 0$, we define the {\bf $s$-distance} $d_s(e,f)$ between two edges $e,f \in E$ as the length of the shortest $s$-walk between them. If there is no $s$-walk between two edges then the $s$-distance is infinite. Aksoy {\it et al.} also  define a number of network science methods generalized from graphs to hypergraphs, including vertex degree, diameter, and clustering coefficients. This work will use their generalization of betweenness centrality and harmonic closeness centrality to hypergraphs using the stratification parameter $s$. 
\begin{itemize}
\item The {\bf $s$-betweenness centrality} of an edge $e$ is
    \[ BC_s(e) := \sum_{f \neq e \neq g \in E} \frac{\sigma^s_{fg}(e)}{\sigma^s_{fg}} \]
where $\sigma^s_{fg}$ is the total number of shortest $s$-walks from edge $f$ to edge $g$ and $\sigma^s_{fg}(e)$ is the number of those shortest $s$-walks that contain edge $e$.

\item The {\bf harmonic $s$-closeness centrality} of an edge $e$ is  the reciprocal of the harmonic mean of all distances from $e$:
    \[ HCC_s(e) := \frac{1}{|E_s|-1} \sum_{\substack{f \in E_s\\ f \neq e}}\frac{1}{d_s(e,f)}\]
where $E_s = \{e \in E : |e| \geq s\}$. We may refer to this as $s$-closeness in this paper, although elsewhere in the literature this term refers to a slightly different concept where the harmonic mean is replaced with the arithmetic mean.
\end{itemize}
Intuitively, harmonic $s$-closeness centrality captures the extent to which a given hyperedge is close in $s$-distance to other hyperedges. In order to have high harmonic $s$-closeness a hyperedge must have small $s$-distance to all (or most) other hyperedges. $s$-Betweenness, on the other hand, identifies bottlenecks in a hypergraph. 
A hyperedge with high $s$-betweenness has many shortest $s$-walks pass through it. 
In comparison, the original formulation of betweenness and harmonic closeness centrality in the setting of graphs has the $s$-distance and number of $s$-paths replaced simply by graph distance and shortest path. 

In order to take into account multiple $s$ values simultaneously in our analysis we average the centrality values across a range of $s$ values and define the \textbf{average $s$-betweenness centrality} and \textbf{average  harmonic $s$-closeness centrality} as
\[ \overline{BC}_s(e) = \frac{1}{s}\sum_{i=1}^s BC_i(e), \qquad 
\overline{HCC}_s(e) = \frac{1}{s}\sum_{i=1}^s HCC_i(e).\]
Computing (average) $s$-centralities for each hyperedge provides a ranked list of hyperedges from most central (high value) to least central.
All hypergraph construction, metric calculations, and  visualizations were performed using the  Python hypergraph library HyperNetX (\url{https://github.com/pnnl/HyperNetX}).

Some conventional approaches to infer graph structures from high-throughput data use correlated gene expression patterns to build connections. In this context, a gene with high degree (i.e. a hub) has similar expression behavior to many other genes, implicating it as a potential master regulator of gene expression. A gene with high betweenness (i.e. a bottleneck) on the other hand, bridges two regions of the graph indicating that it spans two different behavioral profiles. Genes in this position are potentially involved in causing a transition from one response pattern to another. Thus hubs and bottlenecks may represent master gene expression regulators of two different varieties. Previous work by our group and others has shown that graph vertices in hub and bottleneck positions are significantly enriched for genes critical to the process under study \cite{mitchell2019,mitchell2013network,yu2007importance,mcdermott2009bottlenecks}. Given these prior results and biological relevance of centrality in the setting of graphs, we hypothesized that  hypergraph average $s$-betweenness (and potentially average $s$-harmonic closeness) will have similar biological relevance. 

\section{Results}
By analyzing the curated omnibus transcriptomic data set described above from cells infected with five different viruses and their mutants using both graph and hypergraph approaches, we illustrate the advantages of applying a hypergraph approach to uncover the underlying molecular signatures and mechanisms common across host response to viral infection broadly.

\subsection{Hypergraph structure}
We create hypergraphs from transcriptomics $\log_2$-fold change data calculated from gene expression levels of infected experiments relative to time-matched uninfected mock experiments

\begin{figure}[h!]
\centering
\includegraphics[width=0.35\textwidth]{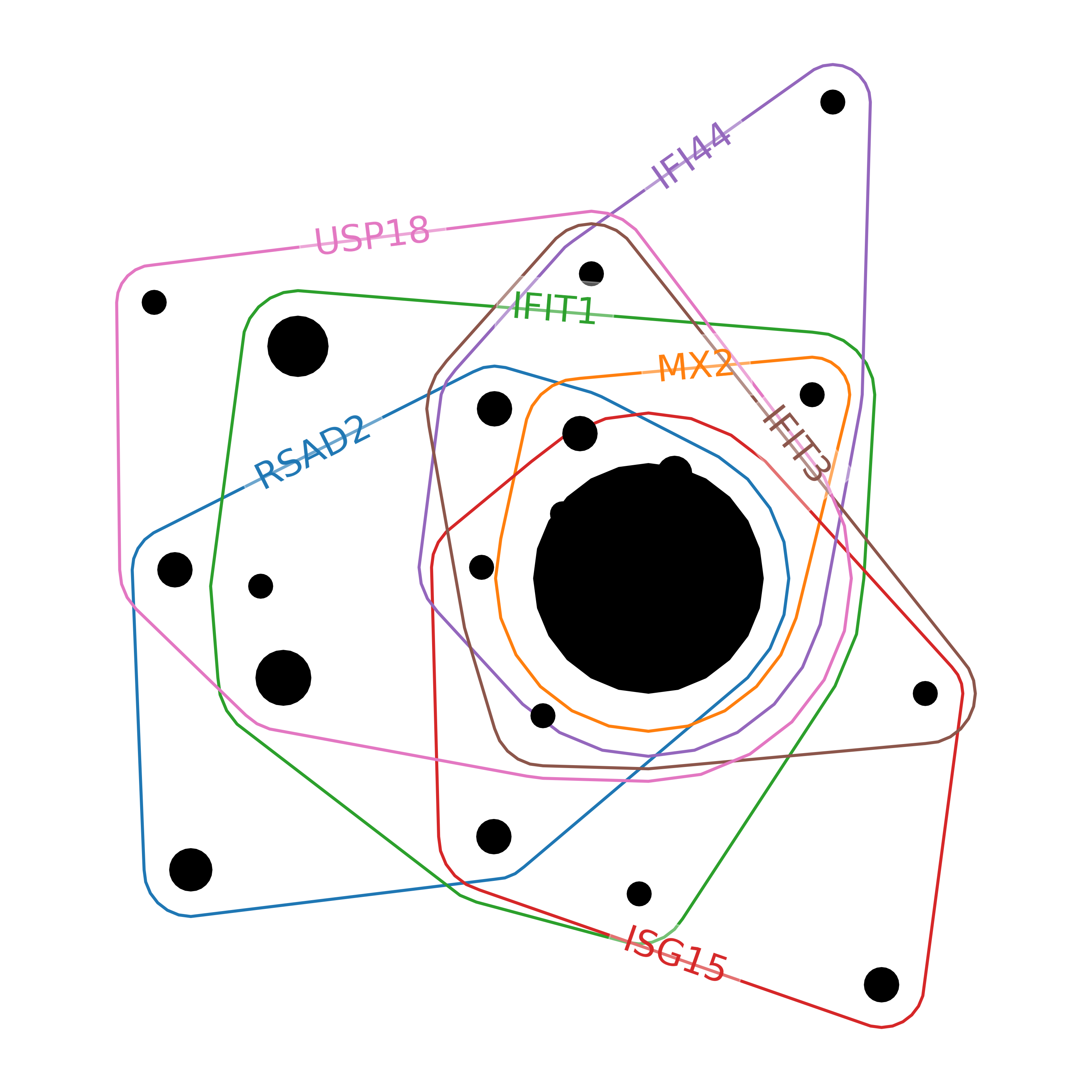}
\caption{Small connected subset of the condition/gene hypergraph. Hyperedges are genes, and black circles indicate groups of vertices (conditions), with larger circles indicating larger groups.}
\label{fig:small_comp_ismb}
\end{figure}

As discussed above, in our hypergraphs hyperedges represent \emph{genes} and vertices represent \emph{conditions}.
The vertex representing condition $X$ is contained in the hyperedge representing gene $G$ if gene $G$ is significantly perturbed, either up- or down-regulated, in condition $X$.
The entire hypergraph represents $n=179$ experimental conditions (vertices) and $m=$7,782 genes (edges).
A small subset of highly connected hyperedges (that is, genes with a large core of common conditions), is shown in Figure \ref{fig:small_comp_ismb}.

\begin{figure}[h!]
\centering
\begin{subfigure}[b]{0.45\textwidth}
    \centering
    \includegraphics[width=\textwidth]{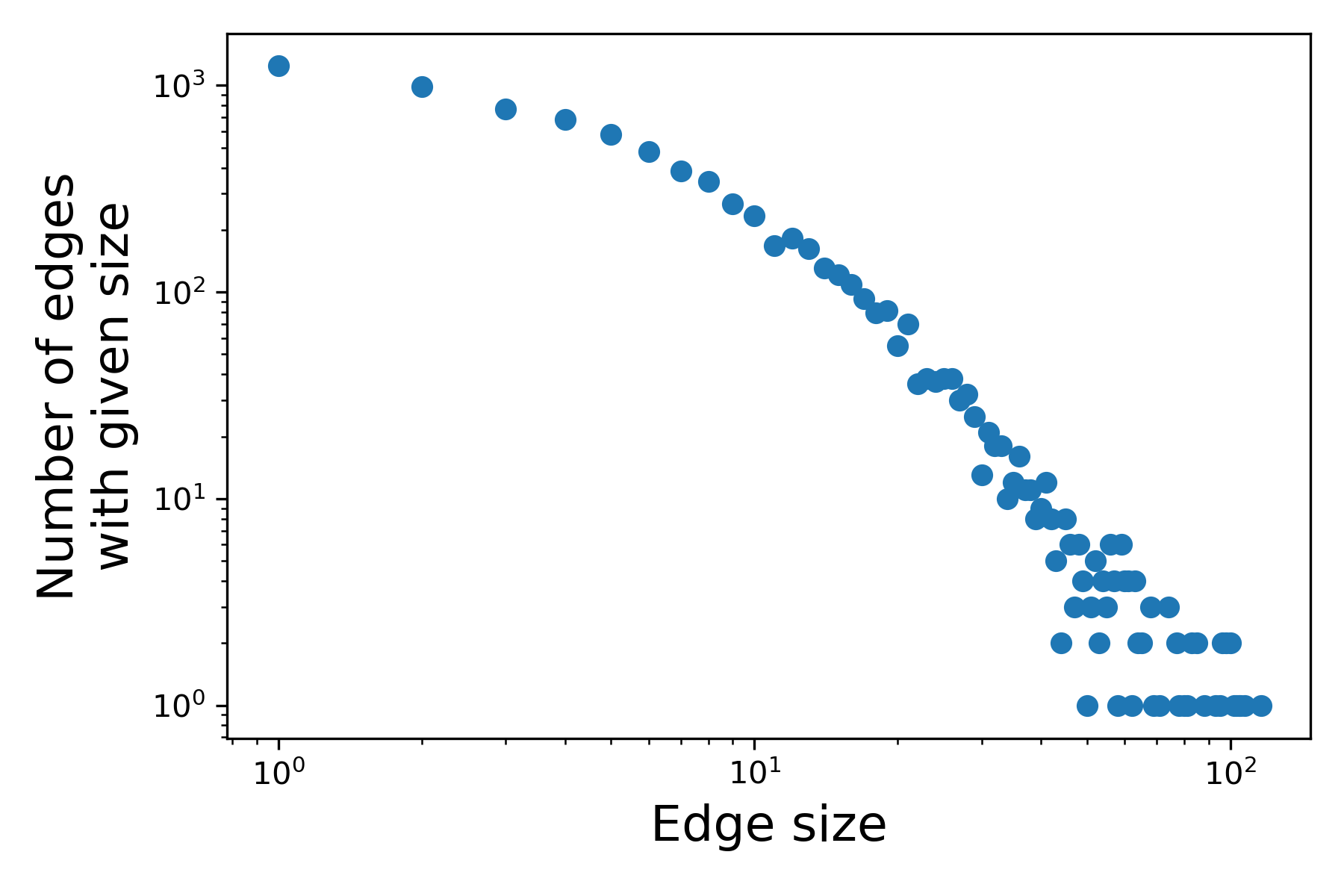}
    \caption{Edge size distribution. Each edge represents a gene, the size of the edge is the number of conditions in which the gene is significantly perturbed.}
    \label{fig:edge_size_dist}
\end{subfigure}
~
\begin{subfigure}[b]{0.45\textwidth}
    \centering
    \includegraphics[width=\textwidth]{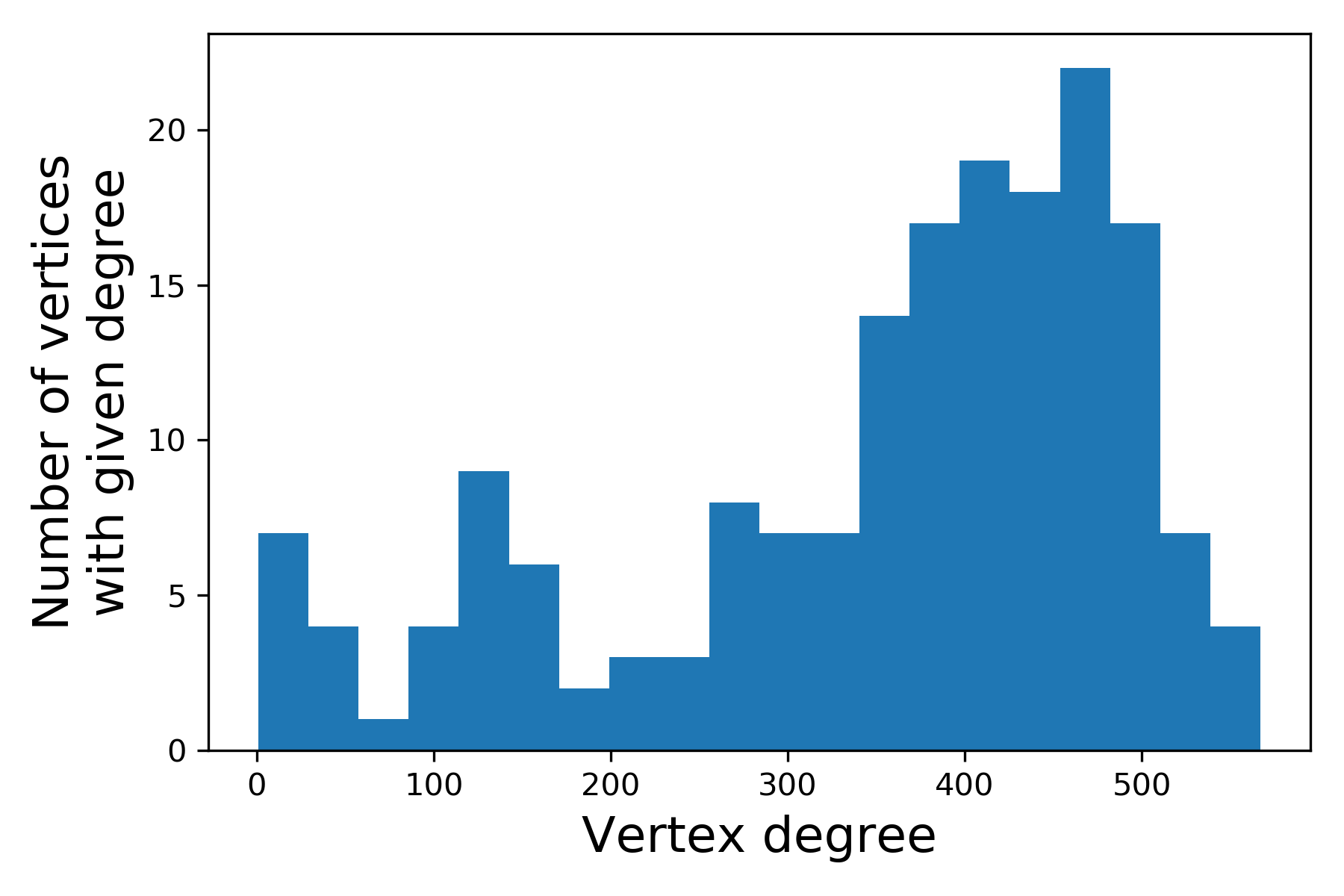}
    \caption{Vertex degree histogram. Each vertex represents a condition, the degree of a vertex is the number of genes significantly perturbed in that condition.}
    \label{fig:deg_dist}
\end{subfigure}
\\
\begin{subfigure}[b]{0.45\textwidth}
    \centering
    \includegraphics[width=\textwidth]{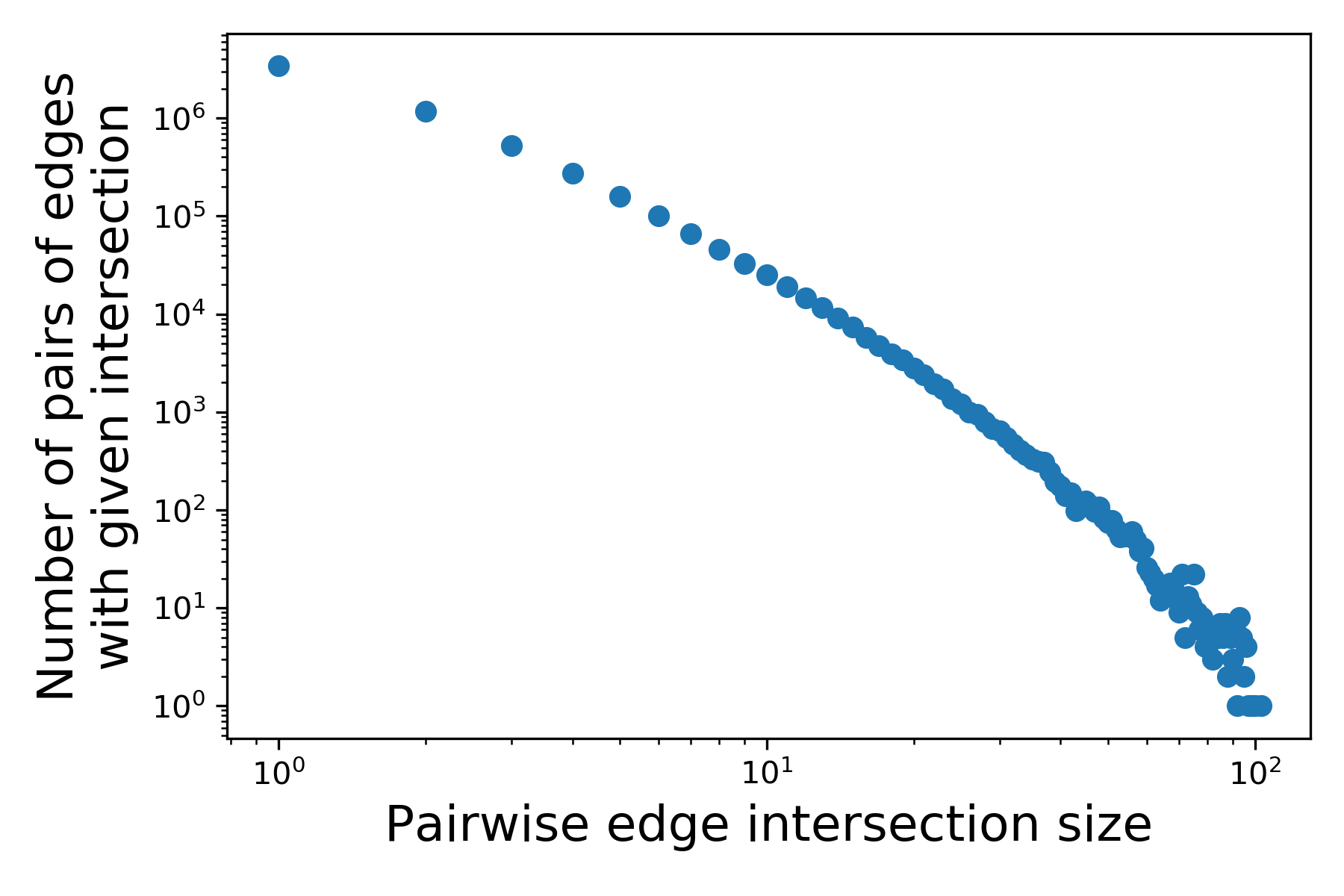}
    \caption{Pairwise edge intersection size distribution. Each edge represents a gene. The intersection between two edges indicates the set of conditions that both genes are significantly perturbed in.}
    \label{fig:inter_size_dist}
\end{subfigure}
\caption{Distributions of simple hypergraph statistics of our hypergraph}
\label{fig:dists}
\end{figure}

Distributions of even fundamental hypergraph statistics can illuminate some of the complex interaction structure present in the data.
Figure \ref{fig:edge_size_dist} shows that the distribution of the sizes of the hyperedges (that is, the number of conditions a gene is significantly perturbed in) is roughly power-law, sometimes referred to as ``heavy tailed''.
This means that there are many genes (1,247 of them) significantly perturbed in only one condition and relatively few genes significantly perturbed in many conditions, with a maximum number of conditions for a single gene on the order of 100.
The six largest hyperedges, with sizes greater than 100 in increasing order, correspond to the genes ISG15, IL6, ATF3, RSAD2, USP18, and IFIT1. All of these genes are part of the interferon response, a critical pathway in response to viral infections \cite{IFIT1McDermott,IFNMenachery}.

On the other hand the vertex degree distribution is the number of edges a particular vertex is contained in (that is, the number of genes significantly perturbed for each condition (vertex)). This is shown as a histogram in Figure \ref{fig:deg_dist}, and it has a very different shape than the edge size distribution.
There are relatively few conditions with small numbers (less than 200) or large numbers (more than 500) of significantly perturbed genes.
The most common number of significantly perturbed genes for a condition is between 400 and 500.
This peak is likely an artifact of how we choose when a vertex is contained within an edge.
The degree of a condition vertex is the number of genes that are significantly perturbed for that condition.
By our procedure these are genes with $z$-score higher than 2 and $p$-value less than 0.05.
If we only used the $z$-score threshold, and our fold change data are normally distributed for each condition, then we would expect that 5\% of the genes would have $z$-score greater than 2.
There are 9,760 genes in our data and 5\% of that  would be 488 genes, which is roughly where the peak is.
The skewness and additional modes of the distribution of degrees is likely due to the addition of the $p$-value condition.

Finally, Figure \ref{fig:inter_size_dist} shows another power-law distribution, this time of the size of pairwise edge intersections, or in other words, the number of conditions that pairs of genes are both significantly perturbed within.
We see that there are many pairs of genes that have few conditions in common and only a few pairs of genes that have many conditions in common, again with a maximum on the order of 100.
The pair of genes with largest intersection is not surprisingly the two largest edges, IFIT1 and USP18, with 103 conditions in common. Interestingly, IFIT1 and USP18 are both well-established interferon response genes, with IFIT1 strongly promoting interferon activity, and USP18 serving to dampen the response \cite{basters2018usp18}.

\subsection{Gene importance rankings}
As noted above, previous studies using graph approaches with similar viral data have demonstrated that network measures like betweenness centrality could be used to identify critical genes \cite{mcdermott2016effect}.
In the present work we hypothesize that extensions of these common graph metrics to a hypergraph, as defined in the Methods section above, can be leveraged to improve upon this prior work.
In particular, we hypothesize that, as has been shown in the graph setting, high $s$-centrality is correlated with the gene being more important in host response to pathogenic viruses.

We calculate average $s$-betweenness centrality, $\overline{BC}_s(g)$, and average harmonic $s$-closeness centrality, $\overline{HCC}_s(g)$, for all genes (hyperedges) in the hypergraph for $1 \leq s \leq 50$.
Both of these $s$-centrality computations provide a numerical value for each gene that can be used to rank the genes from most important (high centrality) to least (low centrality).

To serve as a simpler, but still hypergraph-based, comparison we created another ranked list using hyperedge size for each gene, i.e., the number of conditions the gene was significantly perturbed in.
Larger edge sizes indicate more conditions in which the gene was significantly perturbed and therefore the gene is potentially more important to host response.

In order to compare the centrality rankings from the hypergraph to more common graph methods we employed the CLR methodology that we have used previously to enrich for important genes in a network \cite{mcdermott2016effect,mitchell2019,mitchell2013network}.
The CLR algorithm was run on the matrix of transcriptomics $\log_2$-fold change values using parameters spline = 3 and bins = 10 and the resulting matrix was filtered for all values $\geq 2$.
With this approach, any two genes with shared information above the threshold have similar expression profiles and form a graph edge.
This results in an association graph from which we used the NetworkX graph analytics Python package \cite{hagberg2008exploring} to calculate vertex degree, betweenness centrality, and harmonic closeness centrality.

To provide a simple baseline for comparison a final ranked list was computed directly from the $\log_2$-fold change table without using any graph structure.
For each gene we computed its average absolute value of $\log_2$-fold change and ranked the genes from highest to lowest average.
Higher values mean the gene is more likely to be highly perturbed from the mock-infected samples in many conditions.

In a supplementary file we provide gene rankings for average $s$-betweenness centrality and average harmonic $s$-closeness centrality for $s=50$, hyperedge size, CLR graph betweenness, CLR graph closeness, CLR graph degree, and average fold change.

\subsection{Comparison of rankings}
To ascertain whether our hypergraph rankings are more highly enriched for genes known to be important in host response to viral infection, we gathered three distinct sets of genes:
1) all genes associated with the Gene Ontology (GO) term “immune response” (GO:0006955), downloaded from amigo.geneontology.org, referred to as ‘IR’ hereafter,
2) interferon-stimulated genes gathered from interferome.org (\url{http://www.interferome.org/interferome/search/searchGene.jspx}), referred to as ‘ISG’, and
3) a set of human proteins known to be targets of pathogens acquired from Dyer, et al. \cite{dyer2008}, referred to as ‘PT’.
Although this is a limited set in terms of number of targets, it represents a set collected from a wide number of pathogens, both viral and bacterial, and is a conservative set for assessing the performance of our method and making comparisons between different approaches and parameters.
In Table \ref{tab:gene_sets} we show the size of each gene set (along the diagonal) and the sizes of each pairwise intersection of gene sets (off the diagonal).
Since our data encompasses a wide variety of virus types and infection systems, general immune-related sets were deemed suitable for our purpose.

\begin{table}[h]
    \centering
    \begin{tabular}{l|rrr}
            & IR      & ISG     & PT \\\hline
        IR  & {\bf 1,202}    & 250     & 297 \\
        ISG & 250 & {\bf 1,071}    & 152 \\
        PT  & 297 & 152 & {\bf 906}
    \end{tabular}
    \caption{(Diagonal) Size of each gene set; (Off-diagonal) Size of the  pairwise intersections of the gene sets.}
    \label{tab:gene_sets}
\end{table}

In order to measure the performance of our rankings, we applied gene set enrichment analysis (GSEA) \cite{subramanian2005gene} to each of our gene rankings (average hypergraph $s$-centralities, hyperedge size, CLR centralities, CLR vertex degree, and mean fold-change) using the three immune-related sets as target gene sets.
The GSEA score of a ranked list, computed for a specific gene set, quantifies how concentrated the gene set is at the extremal values of the list.
A high GSEA score means the gene set is concentrated at the top of the list while a low (highly negative) score indicates that the gene set is concentrated towards the bottom of the list.
A score closer to zero means that the gene set is more uniformly distributed throughout the ranked list.
The significance, or $p$-value, of an observed enrichment score, $ES$, is assessed by comparing it with a set of $ES_0$ randomized scores.





\begin{figure*}[h!]
\centering
\includegraphics[width=0.8\textwidth]{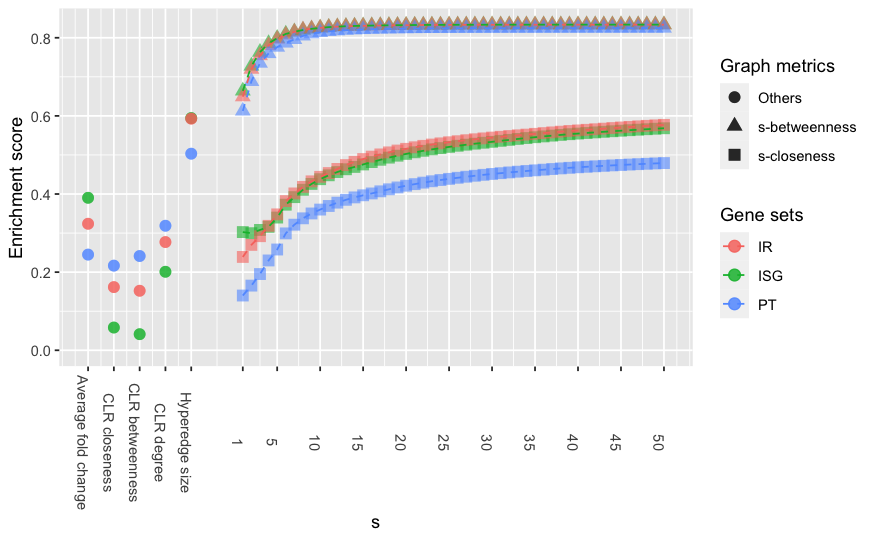}
\caption{Enrichment scores of gene sets using $s$-betweenness and $s$-harmonic closeness metrics. All results are significant with $p < 0.05$.}
\label{fig:GSEA_hypergraphs}
\end{figure*}

Figure \ref{fig:GSEA_hypergraphs} shows GSEA scores for all rankings and for all three target gene sets.
We note the following conclusions: 
\begin{itemize}
    \item Both hypergraph $s$-centrality metrics for most $s$ values, as well as hyperedge size, showed much higher enrichment than lists derived from CLR graphs and average fold change.
    \item But $s$-betweenness enrichment was universally higher than $s$-closeness enrichment, suggesting that these two measurements are capturing fundamentally different behavior within hypergraphs, and that $s$-betweenness appears to be more effective at capturing genes that are important in host responses to viral infection.
\item Both centrality enrichment results (betweeness and closeness) improve significantly when larger $s$ values are taken into account, indicating that when higher order interactions are considered, they  become more powerful in identifying important genes.

\end{itemize}

\begin{figure*}[h!]
\centering
\includegraphics[width=0.75\textwidth]{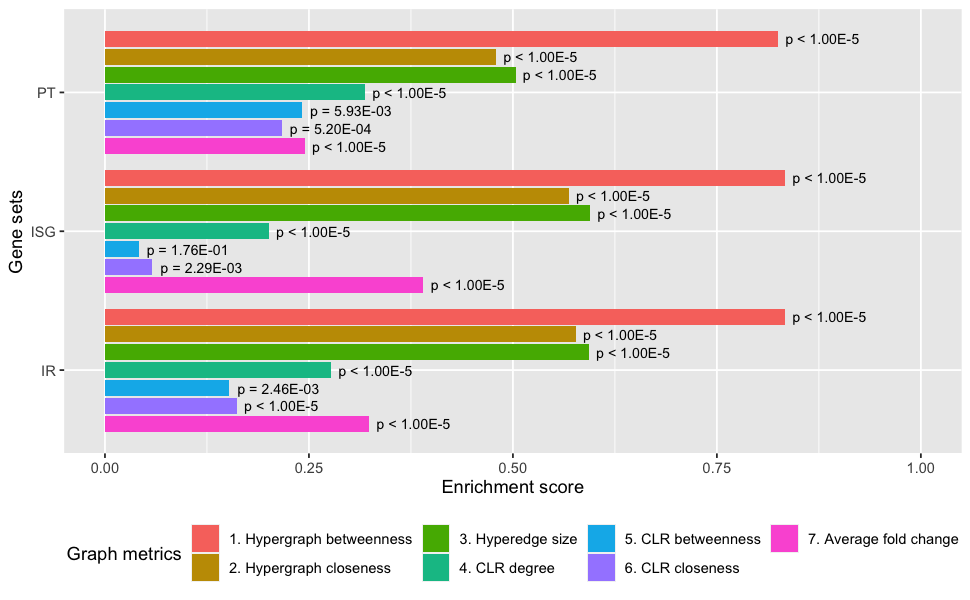}
\caption{Comparison between different hypergraph and graph metrics}
\label{fig:GSEA_comparison}
\end{figure*}

A summary visualization of our results is shown in Figure \ref{fig:GSEA_comparison} taking the rankings for both $s$-betweenness and harmonic $s$-closeness for the highest level of $s=50$ as the representative hypergraph centrality rankings.
We compare those with the five other rankings and again see that $s$-betweenness centrality outperforms all other measures.
While harmonic $s$-closeness centrality outperforms all graph measures it is outperformed by the simple hyperedge size ranking.
The $p$-values, nearly all significantly less than 0.05, are shown in the same plot at the end of the bars.
These results demonstrate that hypergraph $s$-betweenness, but not necessarily harmonic $s$-closeness, considers the complexity of the hypergraph and provides superior performance over graph metrics with regards to identifying biologically important genes.
This aligns with prior work in which betweenness centrality computed for vertices in a graph identifies important genes in a network \cite{mcdermott2016effect,mitchell2019,mitchell2013network,kim2019}.

\section{Discussion}

We draw attention to  two primary observations of interest in our results.
First is the observation that $s$-betweenness centrality consistently outperforms $s$-closeness centrality.
At first glance this seems surprising since betweenness and closeness calculated on the CLR graph have comparable performance.
While the explanation for this result is the subject of ongoing investigation, we observe that these two types of centrality are measuring significantly different properties.
For both graphs and hypergraphs high harmonic closeness centrality indicates that on average a gene is close to many other genes, while high betweenness centrality means that a gene is on many short paths between other genes.
Sometimes these two notions coincide, as seems to be the case in the CLR graph, but there are cases in which they do not.
For example, a gene that may be more on the periphery, i.e., not on many (short) paths, could still be very close to a central core.
Being on few short paths this gene would have very low betweenness.
However, since it is close to a central core it could have high closeness score.

This seems to be the case in the hypergraphs we are studying. 
Since there are many conditions that have a lot of significantly perturbed genes (see Figure \ref{fig:deg_dist}) there is likely a large central core in the hypergraph that increases the closeness scores for peripheral genes, and perhaps all genes.
Indeed we have observed that for small values of $s$ the $s$-closeness values do not correlate with edge size, however for large $s$ values the $s$-closeness scores do tend to correlate with edge size.
This likely means that for low $s$ values the closeness is somehow washed out by this central core and any variability we see is not significant.
In contrast, for $s$-betweenness we see a correlation between edge size and betweenness at all $s$ values.
However, even though $s$-betweenness is correlated to edge size for all $s$ values its enrichment score is still much larger than that for edge size and so seems to be capturing something more significant about hypergraph structure. 

This difference between closeness and betweenness may also be related to the nature of the large gene expression data set used in our study. 
Since both mouse and human-based gene expression data were included in the hypergraph some genes may serve as bridges between different regions of the hypergraph (e.g. predominantly human regions vs predominantly mouse regions).
Genes that are truly important in host response to viral infection would be important across species and more effectively brought to light by the betweenness measure that tends to highlight elements occupying bridge-like positions in the hypergraph. 
Thus, betweenness centrality may be most useful for identifying critical elements when heterogeneous data sets are analyzed.

Our second observation is that our hypergraph $s$-betweenness centrality significantly outperforms established graph centrality techniques. 
This is entirely in keeping with our expectation, as the purpose of  hypergraphs is to capture the complex, multi-way interactions present in a system that are beyond the ability of graphs  to model.
Thus where betweenness centrality has been used in prior studies to identify important biological features the application of hypergraph $s$-betweenness may promote discovery of additional features of interest.
While the finding that hypergraph betweenness represents a new tool for identifying critical hypergraph elements is an exciting contribution of this study, it also presents an additional immediate benefit: genes highly ranked by hypergraph betweenness that do {\em not} appear in any of our target gene sets represent potentially novel discoveries of genes central to viral infection. 
One good example of this is the ZZZ3 gene, which appears in position 4 out of 7,782 in the average hypergraph betweenness ranking, but does not appear in any of the IR, ISG or PT gene sets. 
ZZZ3 is part of the histone reader ATAC complex, which scans the state of histone modification and contributes to gene activation/repression mechanisms \cite{mi2018zz}. 
No known connection between virus infection and ZZZ3 exists, but it may serve a critical role in regulating gene expression in response to general infection. 

Similarly, EPHX1, GDF15, and DUSP1 were not included in the three gene sets and ranked 29, 30 and 33, respectively. These genes are identified as an epoxide detoxification component, a stress responsive cytokine and a stress-responsive phosphatase, respectively. These roles may be related to virus-induced stress in host cells, but the specific mechanisms involved are yet to be elucidated. More exploration of these and other highly ranked genes is the subject of future work for us.

\section{Conclusion}
The work we present in this paper is similar to much of the work surveyed in our literature review in that we show the value of hypergraphs over traditional graph analysis of biological data. 
However, our work differs from these prior studies in a number of ways. First, 
our hypergraphs are built natively from transcriptomics data rather than based on existing graph models of systems.
Although still capturing some multi-way complexities, hypergraphs inferred from graphs may include some induced interactions not actually present in the system that is being modeled. 
Creating hypergraphs natively from the data avoids this imputation.
Other papers we surveyed do create hypergraphs natively from other types of data, but rather than applying centrality measures instead study more structural features like highly connected vertices. 

Previous work \cite{mcdermott2016effect,mitchell2019,mitchell2013network,kim2019} had demonstrated that graph metrics can be used to identify important genes in association graphs, and so we set out to determine if hypergraphs provided an improvement over graphs. 
To assess performance of (hyper)graphs derived from our large viral infection gene expression data set, we identified three gene sets related to virus/pathogen infection and performed an enrichment analysis of our ranked lists compared to these gene sets. 
While the sets were partially overlapping they represented relatively distinct aspects of viral infection in general. 
Our results show that $s$-betweenness, but not necessarily harmonic $s$-closeness, was a useful metric that is able to identify key genes in a comprehensive gene expression data set.
While $s$-closeness does outperform both CLR centrality measures, CLR degree, and average fold change, it does not exceed the performance of a simple ranking according to hyperedge size, which does not require the full hypergraph structure to calculate. 
On the other hand, ranking based on hypergraph $s$-betweenness outperformed all other metrics.

The hypergraphs we created used samples from a wide range of viruses, strains, cell types, and time since infection. 
In future work we plan to apply this measure to compare critical genes in viral response across differing sample features. 
For example, we will split our hypergraph based on pathogenicity (high vs. low), cell type or host, and time since infection (early vs. late). 
Comparing the critical genes across these different hypergraphs may allow us to discover previously unknown indicators of viral infection for early detection or severity determination.
Other future work we plan to pursue includes considering other hypergraph constructions, other data types, and hypergraph algorithms to identify highly connected vertices. 
We plan to combine transcriptomics with proteomics and other `omics measurements to understand whether hybrid hypergraphs yield better results or if the inclusion of more data washes out the complexities.


\section{Back Matter}

\subsection*{Acknowledgements}
The authors thank Dr. Tony Chiang for his helpful comments to improve writing and discussions of biological relevance.

\subsection*{Availability of data and materials}
The datasets generated during and/or analysed during the current study are available in the Gene Expression Omnibus (GEO) repository at the accession IDs listed in the paper, \url{https://www.ncbi.nlm.nih.gov/geo/}.

\subsection*{Funding}
Work in this paper was conducted under the Laboratory Directed Research and Development Program at Pacific Northwest National Laboratory, a multiprogram national laboratory operated by Battelle for the U.S. Department of Energy.
The experimental research and data generation were supported by the National Institute of Allergy and Infectious Diseases under grant number U19AI106772, including an administrative supplement (Ebola) and pilot award (MERS),  and contract number HHSN272200800060C.

\subsection*{Ethics approval and consent to participate}
Human lungs were obtained under protocol 03-1396, which was approved by the University of North Carolina at Chapel Hill Biomedical Institutional Review Board, and donors gave informed consent.

‘Omics data collection studies for animal samples performed at UNC Chapel Hill were performed in animal biosafety level 3 facilities and were conducted under protocols approved by the Institutional Animal Care and Use Committee at UNC Chapel Hill (IACUC protocol \#16-251) according to guidelines set by the Association for the Assessment and Accreditation of Laboratory Animal Care and the U.S. Department of Agriculture.

All animal experiments and procedures performed at UW-Madison were approved by the UW-Madison School of Veterinary Medicine Animal Care and Use Committee under relevant institutional and American Veterinary Association guidelines.

West Nile virus work in mice was carried out in strict accordance with the recommendations in the \emph{Guide for the Care and Use of Laboratory Animals} of the National Institutes of Health. The protocols were approved by the Institutional Animal Care and Use Committee at the Washington University School of Medicine (Assurance number A3381-01).

\subsection*{Author's contributions}
SF (Feng) performed the enrichment analysis, provided biological subject matter expertise, and co-wrote the  manuscript.
BP and EH performed the centrality calculations.
BJ, CJ, and HK developed hypergraph methodology and co-wrote  the manuscript.
JM, and HM provided biological subject matter expertise and co-wrote the  manuscript.
AE, AS, LT, SF (Fan), KW (Walters), PH, DW, QT, VM, TS, AC, JK, KS, NH, LB  contributed to generation, curation, and analysis of data.
MD, RB, KW (Waters), YK supervised generation and analysis of data.
EP developed hypergraph methodology, supervised the hypergraph analysis team, and wrote the manuscript.

\subsection*{Author information}
${}^1$Biological Sciences Division, Pacific Northwest National Laboratory, Richland, WA, USA
${}^2$Department of Mathematics, University of Illinois, Urbana-Champaign, IL, USA
${}^3$Computing and Analytics Division, Pacific Northwest National Laboratory, Seattle, WA, USA
${}^{4}$Systems Science Program, Portland State University, Portland, OR, USA
${}^5$University of Wisconsin-Madison, School of Veterinary Medicine, Department of Pathobiological Sciences, Influenza Research Institute, Madison, WI, USA
${}^{6}$Signature Science and Technology Division, Pacific Northwest National Laboratory, Richland, WA, USA
${}^7$Department of Medicine, Washington University School of Medicine, Saint Louis, MO, USA
${}^8$Department of Epidemiology, University of North Carolina at Chapel Hill, Chapel Hill, NC, USA
${}^{9}$Department of Microbiology \& Immunology, University of Texas Medical Branch, Galveston, TX, USA
${}^{10}$KNOWBIO LLC., Durham, NC, USA
${}^{11}$Department of Pathology \& Immunology, Washington University School of Medicine, St. Louis, MO, USA
${}^{12}$Department of Molecular Microbiology, Washington University School of Medicine, St. Louis, MO, USA
${}^{13}$Department of Comparative Medicine, University of Washington, Seattle, WA, USA
${}^{14}$Division of Virology, Department of Microbiology and Immunology, Institute of Medical Science, University of Tokyo, Tokyo, Japan
${}^{15}$ERATO Infection-Induced Host Responses Project, Saitama, Japan
${}^{16}$Department of Special Pathogens, International Research Center for Infectious Diseases, Institute of Medical Science, University of Tokyo, Tokyo, Japan
${}^{17}$Department of Molecular Microbiology and Immunology, Oregon Health \& Science University, Portland, OR, USA


\bibliographystyle{plain}
\bibliography{hypergraphs_bio}

\end{document}